\documentclass[prl,aps,twocolumn, floatfix,superscriptaddress,showpacs]{revtex4}
\usepackage[naturalnames]{hyperref}
\usepackage{graphicx}
\usepackage{amssymb}
\usepackage{graphicx}
\usepackage{dcolumn}
\usepackage{bm,amsmath,verbatim}
\usepackage{hypcap}
\hypersetup{colorlinks = true, urlcolor = blue, linkcolor = blue, citecolor = blue}

\def\be{\begin{equation}}
\def\ee{\end{equation}}
\def\bea{\begin{eqnarray}}
\def\eea{\end{eqnarray}}
\def\bse{\begin{subequations}}
\def\ese{\end{subequations}}

\renewcommand{\v}[1]{{\bf #1}}

\def\be{\begin{eqnarray}}
\def\ee{\end{eqnarray}}

\newcommand{\expect}[1]{\langle {#1} \rangle}

\makeatother

\begin{document}
\title{Detecting topological superconductivity using low-frequency doubled Shapiro steps}
\author{Jay D. Sau}
\author{F. Setiawan}
\affiliation{Condensed Matter Theory Center and Joint Quantum Institute, Department of Physics, University of Maryland, College Park, Maryland 20742, USA}
\date{\today}
\begin{abstract}
The fractional Josephson effect has been observed in many instances as a signature of a topological superconducting 
state containing zero-energy Majorana modes.  We present a nontopological scenario which can produce a fractional Josephson effect generically in semiconductor-based Josephson junctions, namely, a resonant impurity bound state weakly coupled to a highly transparent channel. We show that the 
fractional ac Josephson effect can be generated by the Landau-Zener processes 
which flip the electron occupancy of the impurity bound state. The Josephson effect signature for Majorana modes become distinct from this 
nontopological scenario only at low frequency.
 We prove that a variant of the fractional ac Josephson effect, namely, 
the low-frequency doubled Shapiro steps, can provide a more reliable signature of the topological superconducting state.
\end{abstract}

\maketitle

Superconductors supporting Majorana zero modes (MZMs)\cite{salomaa,readgreen,kitaev,yakovenko1} at defects 
provide one of the simplest examples of  topological superconductors (TSs) \cite{schnyder,kitaev09}. In fact,
a number of proposals \cite{fu_prl'08,sau,long-PRB,alicea,roman,oreg} to realize such  MZMs have met with considerable success
 \cite{Mou,Den,Das,Churchill,Finck2016,lee2014}. Such systems containing MZMs are particularly interesting~\cite{alicea2012,beenakker2013,leijnse2012,stanescu2013,elliott2015,sarma2015,beenakker2016road}  because of the 
 topologically degenerate Hilbert space and non-Abelian statistics associated with them that  make such MZMs 
useful for  realizing topological quantum computation \cite{nayak_RevModPhys'08}.
While preliminary evidence for MZMs in the form of a zero-bias conductance peak have already been observed  \cite{Mou,Den,Das, Finck2016, yazdani, ucla, shanghai,zhang2016,Churchill,albrecht2016,lee2014}, 
confirmatory signatures of the topological nature of MZMs are still lacking.

The zero-bias conductance peak provides evidence for the existence of zero-energy end modes which can arise not only from TSs but also from a variety of nontopological features associated 
with the details of the end of the system~\cite{Liu,Bagrets,Sau,Brouwer}. In 
contrast, the topological invariant of a TS, being a bulk property, is not affected by the details of the potential at the end. The topological invariant of a one-dimensional TS can be determined from the change in the fermion parity of the 
Josephson junction (JJ)~\cite{kitaev}. Specifically, 
 the  fermion parity of a topological JJ changes when  
the superconducting phase of the left superconductor $\phi$ of the  JJ 
winds adiabatically by $\delta\phi=2\pi$ ~\cite{readgreen,kitaev}.  
 Such a change in fermion parity of the JJ may be detected from the resulting $4\pi$-periodic component
in the current-phase 
relation of the topological JJ~\cite{kitaev,yakovenko}. This is referred to as the fractional Josephson effect 
and can be detected using the fractional ac Josephson effect (FAJE).

 The FAJE involves applying a finite dc voltage $V$ across the junction so that
 the superconducting phase across the junction varies in time as $\phi(t)=\Omega_J t$ \cite{tinkham}.
Here, $\Omega_J=V$ is the Josephson frequency, where we have set $\hbar=1$ and the charge of the 
Cooper pair $2e=1$. The $4\pi$-periodic current-phase relation characteristic of a topological JJ results in a current that has a 
component at half the Josephson frequency, i.e., at $\omega=\Omega_J/2$ instead of $\omega=\Omega_J$ 
characteristic of conventional JJs~\cite{kitaev,yakovenko,fu_kane,roman,oreg,rosenow}. In principle, the resulting 
ac current may be detected by a measurement of the radiation emitted from the junction ~\cite{CPB_LZ,molenkamp2}.
Alternatively, the fractional Josephson effect can also be detected by measuring the size of the voltage steps, known as Shapiro steps~\cite{rokhinson,molenkamp1}. For topological JJs, these voltage 
steps have been numerically found to be $\delta V=2\Omega_J$, which is double the voltage steps for the conventional JJs~\cite{platero,yuri}. 



Interestingly, evidence for both the FAJE~\cite{molenkamp2} and doubled Shapiro steps~\cite{rokhinson,leoshapiro,molenkamp1} 
have been seen in TSs that are expected to support MZMs.  However, there is evidence that such signatures might appear in 
nontopological systems as well.  For example, both the signatures seem to also appear in the TS experiments when the devices are not in the topological 
parameter regime~\cite{molenkamp1,leoshapiro,molenkamp2,Supplementary3}. 
One possible spurious source of FAJE is the period-doubling transition seen in certain JJ systems~\cite{perioddoubling}. In addition, the  FAJE and doubled Shapiro steps 
are known (both experimentally~\cite{CPB_LZ} and theoretically~\cite{sauberghalperin,buttikerfractional}) to arise from Landau-Zener (LZ) processes in certain ranges of frequency.
  Avoiding such LZ processes might require particularly low frequencies in low-noise systems 
 with multiple MZMs~\cite{sticlet}. 
While the LZ process is known to potentially lead to FAJE~\cite{CPB_LZ,sauberghalperin}, there have not been any generic nontopological scenarios presented in the literature so far. 

\begin{figure}
\capstart
\centering
\includegraphics[width=\linewidth]{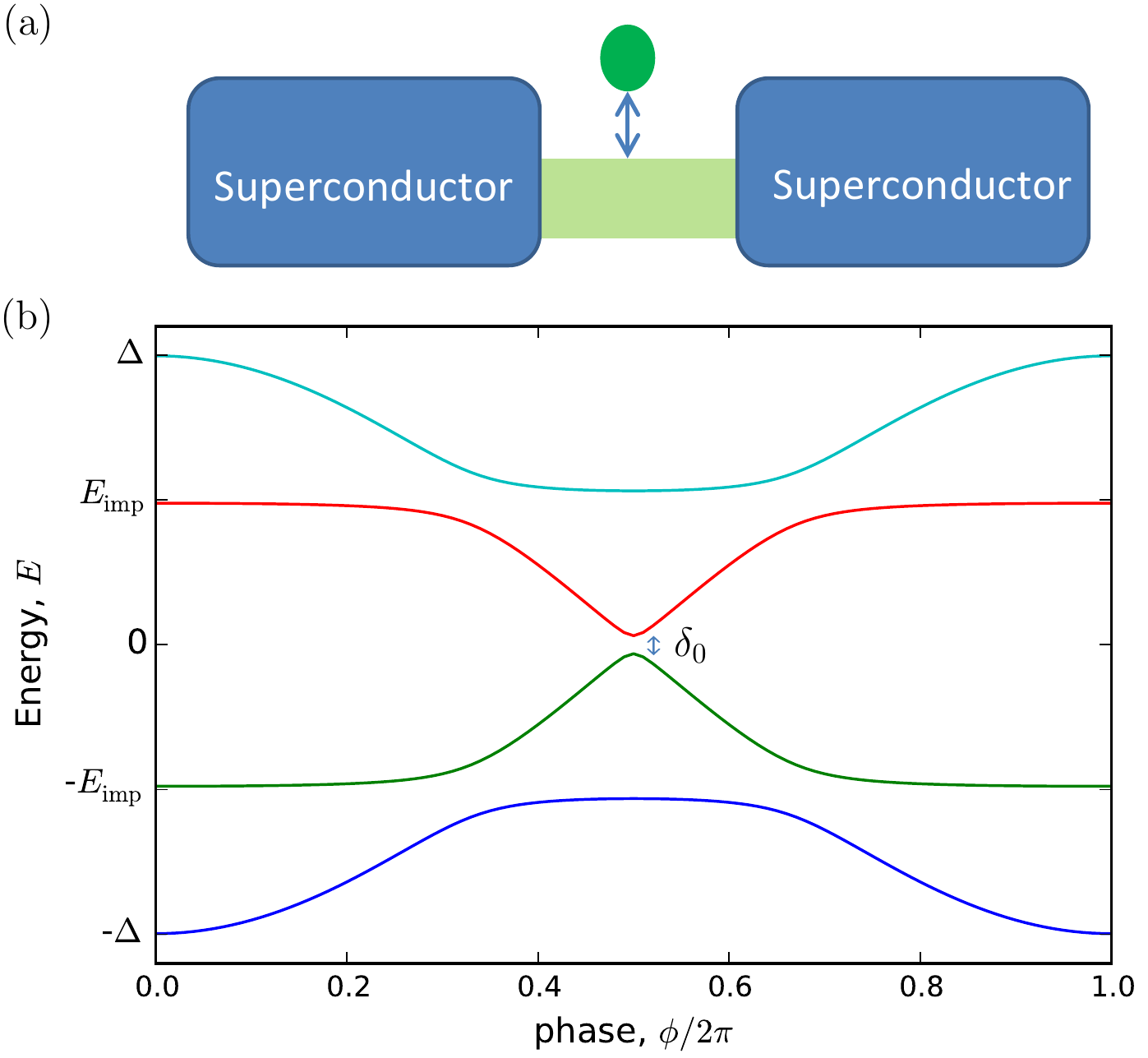}
\caption{(Color online) 
(a) JJ configuration showing FAJE consists of a high transparency channel connecting 
two superconductors. The channel is tunnel coupled to an impurity bound state (shown as a disk adjoining wire).
(b) Computed Andreev bound state spectrum for the setup in (a) shows a weakly avoided crossing 
at $E=0$ and a gap to higher-energy states generated by a larger avoided crossing with the flat impurity bound state.
The weakly avoided crossing can lead to an FAJE at finite voltages.
}\label{Fig1}
\end{figure}

In this Rapid Communication, we start by discussing a generic model of a resonant impurity coupled to a JJ [shown in Fig.~\ref{Fig1}(a)],
which has a weakly avoided crossing in the energy spectrum as a function of phase [see Fig.~\ref{Fig1}(b)]. The present scenario requires only the coexistence of a highly transparent channel in a JJ [as seen in recent measurements of ABS spectra~\cite{vlad}]
 and a weakly coupled impurity bound state. Such a 
coexistence can be found in a multichannel semiconductor-based JJ with a spatially varying density, as is the case of all of the recent 
experiments~\cite{molenkamp1,molenkamp2, leoshapiro,rokhinson}. 
We use a scattering-matrix approach to show that this relatively generic situation can lead to an 
FAJE over a frequency range of a factor of a few even in the absence of any TS.
In order to distinguish between this nontopological scenario from TS, it is important to be 
able to go to ultralow MHz frequencies in the FAJE measurements. Shapiro steps provide the setup where
such a large range of frequencies spanning three orders of magnitudes (MHz--GHz) are possible~\cite{lowShapiro}. 
In the second part of this Rapid Communication, we provide a rigorous framework connecting Shapiro steps to TS where we show that the low-frequency doubled Shapiro steps are guaranteed to appear in the  overdamped driven measurements of topological JJs.

Let us first understand how an FAJE can occur in a nontopological setup such as the setup in
Fig.~\ref{Fig1}(a). For simplicity, we consider the superconductors to be $s$ wave with a highly 
transparent normal channel in between together with a subgap impurity bound state. The highly transparent channel supports Andreev bound states (ABSs) in the junction 
that approach zero energy [see Fig.~\ref{Fig1}(b)] when the phase $\phi$ crosses $\phi=\pi$~\cite{beenakker}. Applying a finite voltage $V$ across the junction causes the superconducting phase $\phi$ to vary in time as $\phi(t)=V t$. This leads to the possibility of LZ processes exciting Cooper pairs across the superconducting gap.  In general, these Cooper pairs are transported across the entire superconducting gap via multiple Andreev reflections~\cite{AverinBardas,klapwijk1982}, ultimately leading to a dissipative but otherwise conventional ac Josephson effect~\cite{AverinBardas}.
This situation is modified when the junction is tunnel coupled to impurity bound states. As shown in Fig.~\ref{Fig1}(b),
the ABS spectrum of the JJ varies with phase $\phi$ where it crosses 
the relatively flat impurity bound state with energy $E_{\mathrm{imp}}$ at pairs of points. At such crossings, the junction exchanges a Cooper pair with 
the flat impurity state. When $\phi = \pi$,
the ABS loses a Cooper pair to the condensate through a LZ process 
across the zero-energy gap $\delta_0$. As the ABS energy approaches the second avoided crossing
with the impurity bound state at energy $E_{\mathrm{imp}}$, the ABS restores its Cooper pair at the expense of leaving the impurity bound state
empty. Thus, the impurity bound state electron occupancy is flipped via the LZ process as the phase varies over a period of $\phi=0$ to $\phi=2\pi$ which is restored during the next $2 \pi$ cycle. Therefore, while the spectrum of the junction is $2\pi$ periodic, the occupation of the impurity bound state is $4\pi$ periodic. Since the total energy $E$ which includes the spectrum and occupation of the ABS and impurity bound states determines the supercurrent $I(\phi)$ by $I(\phi)\sim \partial_\phi E(\phi)$, $I(\phi)$ would also be $4\pi$ periodic with the phase $\phi$. This manifests as a peak in the radiation spectrum from the current at a frequency of $\omega=\Omega_J/2$ instead of the usual Josephson frequency $\omega=\Omega_J$ peak.

\begin{figure}
\capstart
\centering
\includegraphics[width=\linewidth]{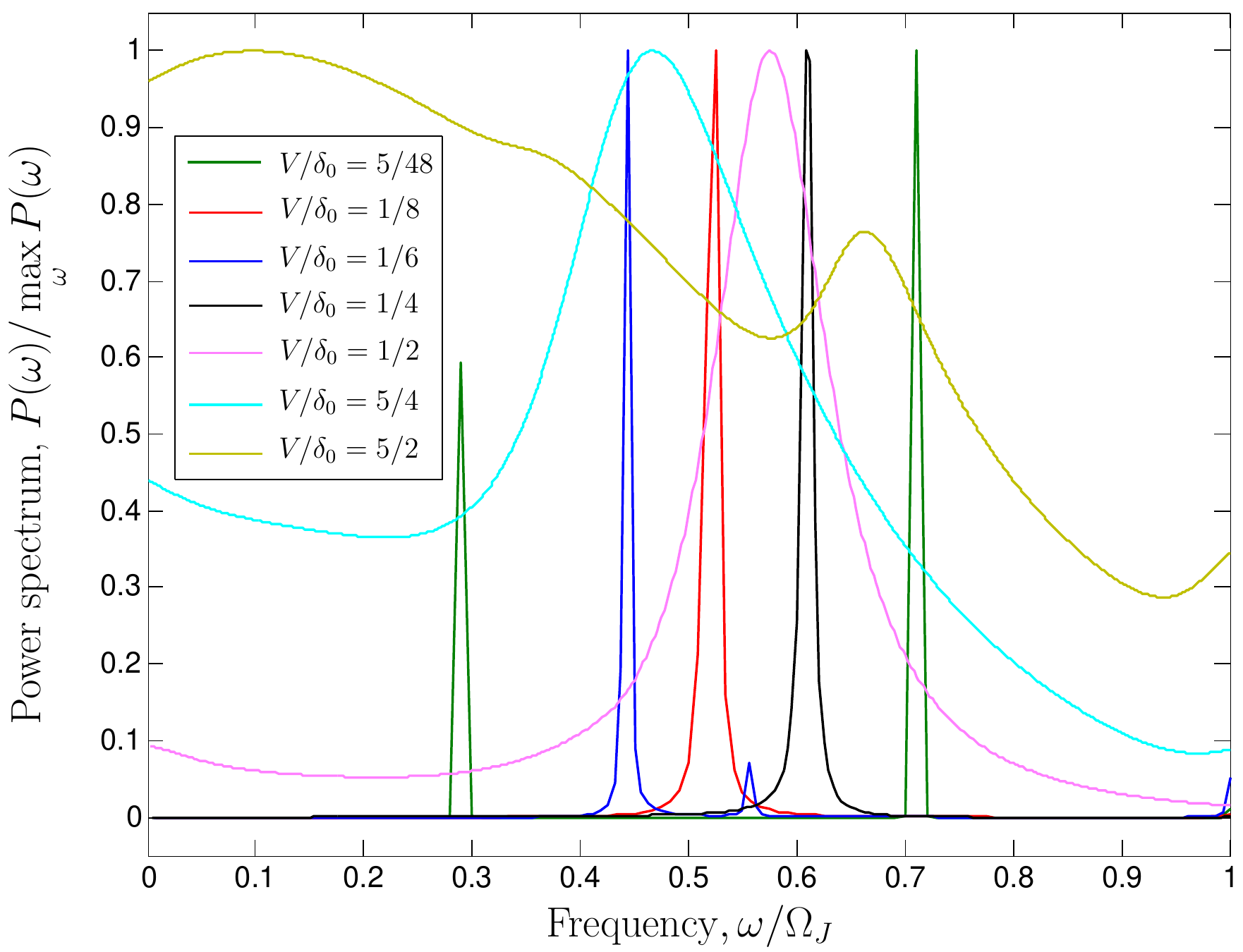}
\caption{(Color online) 
Power radiated $P(\omega)$ as a function of frequency $\omega/\Omega_J$ for
 different ratios of the applied voltage $V$ relative to the zero-energy gap $\delta_0$.
 The power spectrum $P(\omega)$ shows a fractional ac Josephson peak at $\omega= \Omega_J/2$ for 
a range of values of  $V/\delta_0$. The peak broadens out at higher voltages and shifts towards a more 
conventional peak at $\omega=\Omega_J$ at lower frequency (while becoming smaller).  $P(\omega)$ 
has been rescaled so that all peaks are clearly visible.
}\label{Fig2}
\end{figure}

While the qualitative argument above suggests the possibility of an FAJE occurring in nontopological semiconductor systems, it assumes the zero-energy LZ processes to be perfect and all other LZ processes to be completely avoided. In the following, we perform a completely unbiased quantitative analysis of the FAJE for the JJ shown in Fig.~\ref{Fig1}.
To begin with, we note that at any finite voltage $V$, the occupation of an ABS  fluctuates due to excitations out of the bulk gap (via multiple Andreev reflections). The quasiparticle fluctuations ensure that the system 
equilibrates to the grand canonical ensemble (with no conserved fermion parity) such that the expectation value of the current is $2\pi$ periodic 
as in the conventional system ~\cite{bloch}. Thus, strictly speaking, the FAJE at any finite voltage is subject 
to random fluctuations and can only appear in the noise spectrum of the current~\cite{meyer,badiane2013ac}. 
To assess the range of voltages over which the JJ shown in Fig.~\ref{Fig1}(a) exhibits an FAJE, we compute the 
noise spectrum of the current 
\begin{equation}
P(\omega)=\int d\tau e^{i\omega \tau}[\expect{I(t)I(t+\tau)}-\expect{I(t)}\expect{I(t+\tau)}],
\end{equation}
where $\expect{\cdots}$ denotes the averaging over time $t$. The current~\cite{AverinBardas} and its noise spectrum~\cite{meyer,badiane2013ac} can be computed by considering the scattering of quasiparticles between the superconducting leads,
which are at different voltages. This approach has the advantage of including the contribution 
 of not only the low-energy ABSs but also all bound and scattering states in the junction. We have expanded this formalism 
to general superconductor-normal-superconductor junctions~\cite{Supplementary1}. Our general framework can be easily implemented with Kwant~\cite{KWANT} which supplies the normal-superconductor scattering matrices.
The resulting power spectrum $P(\omega)$ is plotted against the frequency scaled by the Josephson frequency, i.e., $\omega/\Omega_J$ in Fig.~\ref{Fig2}
for various voltages for the system depicted in Fig.~\ref{Fig1}(a) with the spectrum shown in Fig.~\ref{Fig1}(b).
 The power spectrum at high voltages is quite broad, which becomes narrower at lower frequency and develops peaks in the vicinity of 
$\omega/\Omega_J=1/2$ before splitting off to different values. The high-frequency spectrum is also several orders smaller in magnitude, which is 
expected in the adiabatic limit when fluctuations in the ABS occupation are small. While some of the peaks appear to move away from the 
ideal fractional value and come back, this might be difficult to resolve at a high level of broadening arising from nearby energy states 
and circuit-noise induced broadening.

The spurious FAJE peaks in Fig.~\ref{Fig2} resulting from the LZ mechanism appear over a frequency range narrower
compared to the parametrically large frequency range (i.e., $ \Gamma, \delta \leq\omega \leq \Delta$) of the FAJE in a high-quality TS~\cite{nazarov, aguado, meyer,badiane2013ac}. Here, {\color{blue}$\Delta$} is the induced superconducting gap, which is a relatively large frequency ($\sim$ GHz), and $\Gamma$
and $\delta$ are respectively
 the quasiparticle poisoning rate and the MZM overlap that become vanishingly small ($\lesssim$ MHz) in high-quality TSs.

It is clear from Fig.~\ref{Fig2} that distinguishing a bona fide TS from an LZ-type mechanism induced by resonant bound states
requires  low-frequency ($\lesssim 50$ MHz) measurements of high-quality TS devices  with $\Delta \gg \delta,\Gamma$. The FAJE which involves measuring small oscillating currents is difficult to perform for low frequencies because such small oscillating currents are typically 
measured using on-chip detectors~\cite{CPB_LZ,Kouwenhoven03}  that are suited to measure 
relatively high frequencies ($\sim$ GHz). On the other hand, the Shapiro step~\cite{tinkham}, which is a variant of the FAJE, 
has been demonstrated over a large range of frequencies from several MHz to GHz~\cite{lowShapiro}. While this makes the Shapiro step 
promising for the detection of TSs, a rigorous proof establishing the doubled Shapiro step as a signature of TS is still missing from the literature. Below, we demonstrate analytically that the low-frequency doubled Shapiro steps can be used as a reliable signature of TS.

We begin by considering the Shapiro step experiment where a JJ shunted with a resistance $R$ is biased with a time-varying current $I_{\mathrm{bias}}(t)=I_{\mathrm{dc}}+I_{\mathrm{ac}}\cos{(\Omega_J t)}$, 
with $I_{\mathrm{dc}}$ and $I_{\mathrm{ac}}$ being dc and ac bias currents, respectively. For the following analysis, we make a \textit{key assumption} that we are working in the limit of low-frequency $\Omega_J$ so that the Josephson current 
$I_J(\phi(t))$ can be taken to be in equilibrium, apart from the conserved local fermion parity. 
The assumption of being at sufficiently low frequency can only be justified by studying the Shapiro steps over a few orders of magnitude in frequency (from $\Delta \sim$ GHz to $\delta, \Gamma \sim$ MHz). Using this assumption and the result of Bloch~\cite{bloch}, 
we can establish that $I_J(\phi)$ for any nontopological system must be $2\pi$ periodic and thus rule out any nontopological FAJE such as those from 
the LZ mechanism.

Furthermore, assuming that the shunt resistance $R$ is small enough to allow the JJ to be overdamped, the 
equation of motion for $\phi(t)$ for the resistively shunted JJ takes the standard form~\cite{tinkham}  
\begin{equation}
\frac{d\phi}{dt}=R[I_{\mathrm{bias}}(t)-I_J(\phi(t))].\label{Shap_eom}
\end{equation}
For illustration purposes, we will choose a simple case of $I_J(\phi)=I_0\cos{(2\pi\phi)}+I_{\mathrm{top}}\cos{(\pi\phi)}$, where $I_0$ and $I_{\mathrm{top}}$ are the 2$\pi$- and 4$\pi$-periodic components of the critical current of the adiabatic current-phase relation, respectively. However, our results generally hold and do not depend on this 
parameter choice as is proven by the analytic arguments in Ref.~\cite{Supplementary2}.
The dc voltage $V$ across the JJ  is calculated by considering the average change of the phase 
\begin{equation}
V=\lim_{t\rightarrow \infty}\frac{\phi(t)-\phi(0)}{t},\label{Vdc}
\end{equation}
where the limit is computed by choosing a sufficiently long simulation time for Eq.~\ref{Shap_eom}. 

\begin{figure}
\capstart
\centering
\includegraphics[width=\linewidth]{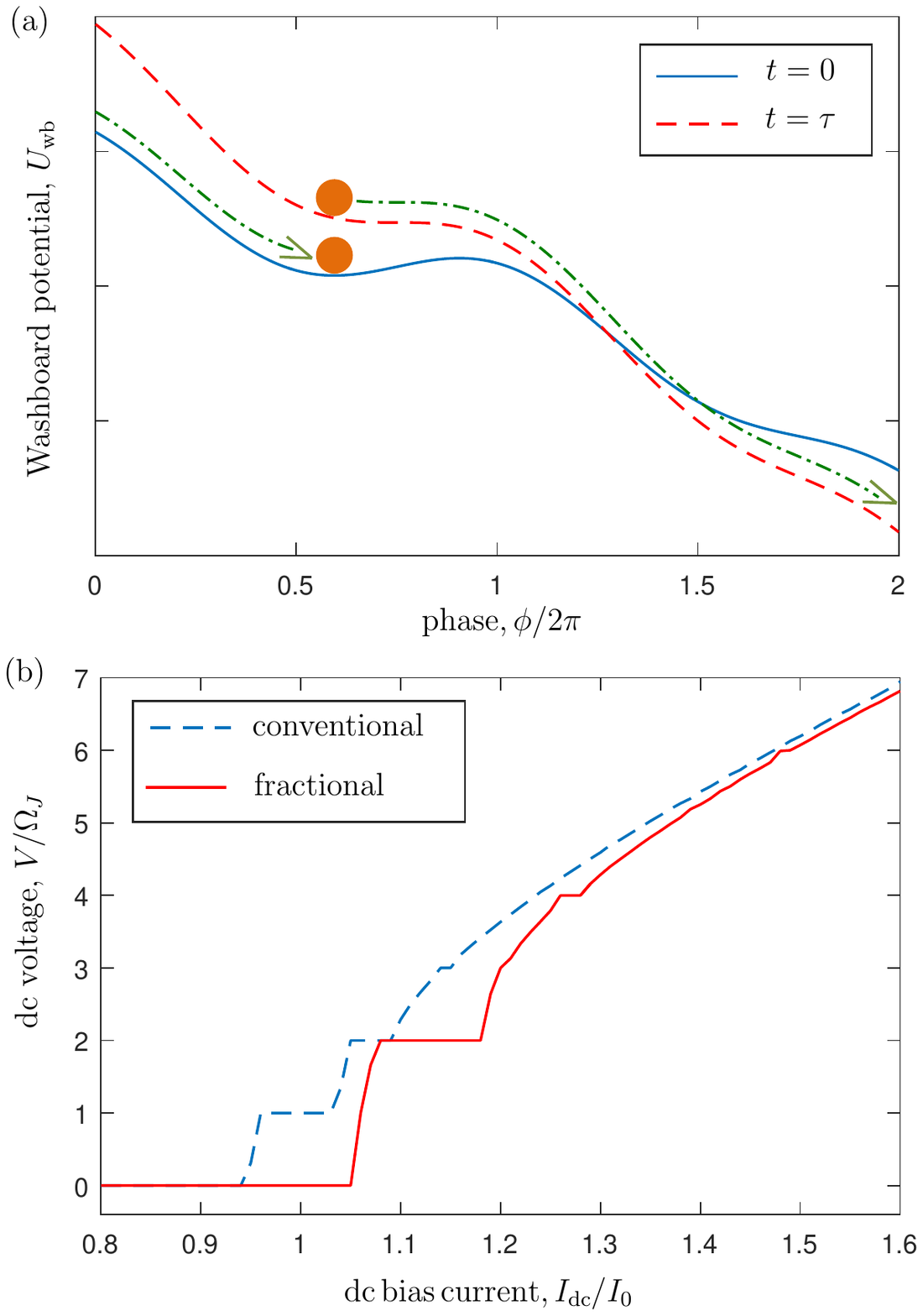}
\caption{(Color online) 
(a) Schematic of a phase particle (orange disk) on a tilted washboard potential that describes the phase dynamics in an overdamped JJ.
As the bias current increases from $t=0$ to $t=\tau$, the phase particle is released from the local minimum and traverses the trajectory along the green dashed-dotted 
arrow,
 and stops when the current bias is back to its value at $t=0$ and the phase 
particle has traveled by $4\pi$  (for the TS case shown here). This corresponds to a voltage step of 2$\Omega_J$. (b) Shapiro step calculated numerically for a putative fractional Josephson system
shows doubled Shapiro steps (see also Ref.~\onlinecite{platero}) as opposed to a conventional system with all integer Shapiro steps for an overdamped JJ.
 Here, $I_{\mathrm{ac}}=0.1I_0$, $R=25$, $I_{\mathrm{top}}=0.15I_0$ (for fractional), and $I_{\mathrm{top}}=0$ (for conventional).
}\label{Fig3}
\end{figure}

We will now show that overdamped JJs constructed out of TSs are generically characterized by a doubled Shapiro step in the strongly overdamped 
and low-frequency limit (i.e., $\Omega_J/I_J R\ll 1$). 
The dynamics of $\phi(t)$ described by Eq.~\ref{Shap_eom} can be 
understood simply by an analogy of a ``phase particle" rolling down a washboard potential according to the equation
 $\dot{\phi}(t)=-\partial_\phi U_{\mathrm{wb}}(\phi,t)$, where the washboard potential is written as $U_{\mathrm{wb}}=-R[I_{\mathrm{bias}}(t)\phi-\int d\phi I_J(\phi)]$. 
As seen in Fig.~\ref{Fig3}(a), 
because of the ac drive, the potential $U_{\mathrm{wb}} (\phi,t)$ varies in time with local minima at each cycle when
 $\phi(t)=\phi_0$
such that
\begin{align}
&I_{\mathrm{bias}}(t)-I_J(\phi_0)=0.
\end{align}
In the adiabatic limit (i.e., $\Omega_J/I_J R\ll 1$), one can show that the phase particle approaches the minimum of the washboard potential exponentially in time
once every period of the drive. This leads to a well-defined voltage that appears as a sharp plateau in the Shapiro steps~\cite{Supplementary2}.

Let us for now assume that~\cite{Supplementary2} the phase particle approaches a minimum of $U_{\mathrm{wb}}$ during the time interval 
when such exists.  In the conventional 
case of a $2\pi$-periodic function $I_J$, this can occur once in a $2\pi$ period provided the critical current $I_{J,\mathrm{max}}>(I_{\mathrm{dc}}-I_{\mathrm{ac}})$.
This will certainly occur if $I_{\mathrm{dc}}$ is small enough. In addition, if  $I_{\mathrm{dc}}>(I_{J,\mathrm{max}}-I_{\mathrm{ac}})$, then there will be a range of time when $U_{\mathrm{wb}}$ 
has no minimum and the adiabatic solution breaks down. In this case, $\phi(t)$ will wind by a multiple of $2\pi$ and collapse to $\phi_0$ after 
a winding of $2\pi n$. 
The result is that an integer voltage appears across the JJ.
In the case of a topological JJ, the current-phase relation $I_J(\phi)$ has a $4\pi$-periodic component and one can define two critical currents $I_{J,\mathrm{max}}$ and $I_{J,\mathrm{max}}'$, one associated 
with the range $\phi\in [4n\pi,(4n+2)\pi]$ and the other in the range $\phi\in [(4n-2)\pi,4n\pi]$. In our simple model $I_{J,\mathrm{max}},I_{J,\mathrm{max}}'=I_0\pm I_{\mathrm{top}}$. 
As in the conventional case, the dc bias current must satisfy 
$I_{\mathrm{dc}}>(I_{J,\mathrm{max}}-I_{\mathrm{ac}})$ (assuming $I_{J,\mathrm{max}}>I_{J,\mathrm{max}}'$) to exit the zero-voltage state even in the TS case. On the other hand, if $2I_{\mathrm{ac}}<(I_{J,\mathrm{max}}-I_{J,\mathrm{max}}')$, 
then $I_{\mathrm{dc}}>I_{J,\mathrm{max}}'+I_{\mathrm{ac}}$ so that the phase particle cannot stop at one half of the minima. This leads to a doubled voltage 
step for the topological case, as seen from the numerical solution of Eq.~\ref{Shap_eom} [see Fig.~\ref{Fig3}(b)].

In summary, we have shown that while the FAJE can be viewed as a smoking gun for the TS with MZMs, 
a detailed study of the frequency dependence of the FAJE is necessary before concluding a system to 
have realized the TS. We have shown this by considering a  generic model of a high transparency channel in a JJ 
coupled weakly to a resonant impurity. We find  this model to show an FAJE quite generically in semiconductor-based JJs, similar to the TS case with MZMs. Nevertheless, TSs are expected to show FAJE over a parameterically larger range of frequency. We argue that the current-phase relation over such a range of frequency, particularly 
at the low-frequency end, is better studied by considering the Shapiro step experiment. We present a way of understanding the 
Shapiro step experiment in terms of the tilted washboard potential that guarantees that the necessary and sufficient condition for the existence of 
doubled Shapiro steps in the low-frequency limit is that the JJ is formed from a TS. Thus, low-frequency Shapiro steps which have 
been demonstrated in conventional systems can serve as a smoking gun for MZMs.

This work is supported by Microsoft Station Q, Sloan Research Fellowship, NSF-DMR-1555135 (CAREER), and JQI-NSF-PFC. We acknowledge enlightening discussions with Anton Akhmerov, Julia Meyer, Leo Kouwenhoven, 
Attila Geresdi, Yuli Nazarov, Chang-Yu Hou, Sergey Frolov, Ramon Aguado,  
and Roman Lutchyn in the course of this work.
J.D.S. is grateful to the Aspen Center for Physics where part of this work 
was completed.

\bibliographystyle{apsrev4-1}

\begin{thebibliography}{99}
\bibitem [{Sup({\natexlab{a}})}]{Supplementary3}%
  \BibitemOpen
  \href@noop {} {} \bibinfo {note} {While these systems have
  been shown to be topological for the correct gate voltages, the fractional
  Josephson signature appears also for gate voltages where the conductance is
  not at the topologically quantized value.  Disorder scattering between the implied additional 
modes and the topological edge modes ultimately limit the topological robustness of the system to a short
  time scale.}\BibitemShut {Stop}%
\end{thebibliography}

%

\vspace{1cm}
\begin{center}
{\bf\large Supplemental Material for ``Detecting topological superconductivity using low-frequency doubled Shapiro steps"}
\end{center}
\vspace{0.5cm}

\setcounter{secnumdepth}{3}
\setcounter{equation}{0}
\setcounter{figure}{0}
\renewcommand{\theequation}{S-\arabic{equation}}
\renewcommand{\thefigure}{S\arabic{figure}}
\renewcommand\figurename{Supplementary Figure}
\renewcommand\tablename{Supplementary Table}
\newcommand\Scite[1]{[S\citealp{#1}]}
\newcommand\Scit[1]{S\citealp{#1}}

\makeatletter \renewcommand\@biblabel[1]{[S#1]} \makeatother
\section{Calculation of the power spectrum for the fractional ac Josephson effect}
The current noise can be calculated using the scattering matrix approach similar to Refs.~\cite{meyer,badiane2013ac}. Below we discuss a generalization of this approach that 
allows us to calculate the noise numerically in the general case. The scattering matrix is described in terms of current amplitudes $J^{\rho}_{\ell}$ 
where $\rho=\pm$ represents the left and right movers, $\ell=L,R$ denotes the left and right superconductors and $\ell=NL,NR$ are the left- and right-half of the normal 
intervening region in between the two superconductors. This region is infinitesimally small and only there to allow computation of the scattering matrices 
$S_L$ and $S_R$ of the left and right superconductors. 
In terms of these amplitudes, the scattering matrix equations at the interfaces $L\rightarrow NL$, $NL\rightarrow NR$ and $NR\rightarrow R$ are 
written as  
\begin{subequations}
\begin{align}\label{eq:matchingeq}
\left( \begin{array}{c}
\mathcal{J}_{L}^{-,\gamma} (E_n) \\
\mathcal{J}_{NL}^{+,\gamma} (E_n)
\end{array}\right) &= S_L(E_n) \left(\begin{array}{c} \mathcal{J}_{L}^{+,\gamma} (E_n) \delta_{n,0}\delta_{\gamma,L} \\\mathcal{J}_{NL}^{-,\gamma} (E_n) \end{array} \right) ,\\
\left( \begin{array}{c}
\mathcal{J}_{NL}^{-,\gamma} (E_n) \\
\mathcal{J}_{NR}^{+,\gamma} (E_n) 
\end{array}\right) &= \sum_{n'} S_N(E_n,E_{n'}) \left( \begin{array}{c} \mathcal{J}_{NL}^{+,\gamma} (E_{n'}) \\ \mathcal{J}_{NR}^{-,\gamma} (E_{n'}) \end{array}\right),\\
\left( \begin{array}{c}
\mathcal{J}_{R}^{+,\gamma} (E_n)\\
\mathcal{J}_{NR}^{-,\gamma} (E_n) 
\end{array}\right) &= S_R(E_n) \left(\begin{array}{c} \mathcal{J}_{R}^{-,\gamma} (E_n) \delta_{n,0} \delta_{\gamma,R} \\ \mathcal{J}_{NR}^{+,\gamma} (E_n) \end{array} \right),
\end{align}
\end{subequations}
where the superscript $\gamma = L/R$ denotes whether the incoming current is from the left/right superconductor, $E_n = E+ nV/2$ with $n$ being an integer and $E$ being the incoming quasiparticle energy (as in the main text, we set $2e = 1$), and $\mathcal{J}_{\ell}^{\rho,\gamma} = (j_{\ell}^{e,\uparrow,\eta,\gamma},j_{\ell}^{e,\downarrow,\eta,\gamma},j_{\ell}^{h,\uparrow,\eta,\gamma},j_{\ell}^{h,\downarrow,\eta,\gamma})^{\mathrm{T}}$ is the current amplitude in the particle-hole space.

The normal-state transmission is perfect up to a cutoff after which it vanishes completely. The transmission part of the scattering matrix $S_N$ is written as 
\begin{align}
&t_N(E_n,E_{n'})=t_n\left[\frac{1+\tau_z}{2}\delta_{n,n'-1}+\frac{1-\tau_z}{2}\delta_{n,n'+1}\right],
\end{align}
where $\tau_z$ is the $z$-Pauli matrix in the particle-hole subspace, $t_n=1$ in the transmitting energy interval and zero elsewhere.
The reflecting part of $S_N$ is analogously defined as 
 \begin{align}
&r_N(E_n,E_{n'})=r_n\left[\frac{1+\tau_z}{2}\delta_{n,n'}+\frac{1-\tau_z}{2}\delta_{n+1,n'+1}\right],
\end{align}
with $r_n=\sqrt{1-t_n^2}$.

The current noise spectrum $P(\omega)$ can be written in terms of the eigenstates of the system as 
\begin{equation}
P(\omega)=\sum_n |\langle 0|J(\omega)|n\rangle|^2,
\end{equation}
where $|0\rangle$ is the state with all incoming quasiparticles from the occupied bands of the superconductors and $|n\rangle=c_a c_b|0\rangle$ are 
excited states with  negative-energy quasiparticle states $c_a$ and $c_b$ having been emptied.  We will assume that the frequency $\omega$ in the 
current operator is smaller than the Josephson frequency so that $\omega<V$. Furthermore, we will assume that the chemical potential in the normal region 
is very large so that we can assume the group velocity to be constant. With these approximations, the current operator $J(\omega)$ (as a matrix in the current 
amplitude basis) is written as 
\begin{align}
&J(\omega)= 2\pi\eta_z\tau_z \delta(E_{n,a}+E_{n',b}-\omega),
\end{align}
where $\eta_z$ is the $z$-Pauli matrix in the left- or right-mover subspace, $E_{n,a}=E_a+n V/2$ and $E_{n'b}=E_b+n'V/2$ and $E_{a,b}<0$ are quasiparticle energies. Flipping the energies of one of the states by a particle-hole transformation, we have
\begin{align}
P(\omega)\sim \int_{E_a<0,E_b>0}& dE_a dE_b \sum_{\gamma=L/R} |\langle \mathcal{J}^\gamma_{NL,a}|\eta_z\tau_z|\mathcal{J}^\gamma_{NL,b}\rangle|^2 \nonumber\\
& \times\sum_n\delta(E_a-E_b+ nV/2-\omega),
\end{align} 
where $\mathcal{J}^\gamma_{\mathrm{NL}} = (\mathcal{J}_{\mathrm{NL}}^{+,\gamma},\mathcal{J}_{\mathrm{NL}}^{-,\gamma})^\mathrm{T}$.
\section{Analysis of  adiabatic Shapiro step equation}
The goal of this section is to develop an analytic understanding of the Shapiro step equation with the end goal 
of proving the doubling of the Shapiro step period in the topological case. We start with the basic equation of an 
overdamped Josephson junction, which is justified at sufficiently low frequencies, i.e.,
\begin{equation}
\frac{d\phi}{dt}=R[I_{\mathrm{bias}}(t\Omega_J)-I_J(\phi(t))],\label{Shapeom1}
\end{equation}
where $R$ is the circuit resistance and $\Omega_J$ parametrizes the frequency of the drive. 
We make no assumptions on the specific form of either $I_{\mathrm{bias}}$ or $I_J$ other than that they are periodic and 
the equation $I_{\mathrm{bias}}=I_J$ has a solution for most of the time interval (in a sense to be made precise later). In the 
tilted washboard picture where the washboard potential is defined as 
\begin{equation}
\partial_\phi U_{\mathrm{wb}}=R[I_J(\phi)-I_{\mathrm{bias}}(t\Omega_J)],
\end{equation}
this is equivalent to requiring that the washboard potential has local minima for most of the time.

By rescaling time variable as $t\rightarrow t R^{-1}$, we can write the equation of motion [Eq.~\eqref{Shapeom1}]
as
\begin{equation}\label{eq:Shapeom2}
\frac{d\phi}{dt}=[I_{\mathrm{bias}}(t R^{-1}\Omega_J)-I_J(\phi(t))].
\end{equation}
We note that in the adiabatic limit ($\Omega_J R\rightarrow 0$), the current bias $I_{\mathrm{bias}}$ in the vicinity of some time $t\sim \tilde{t}$, can be 
approximated to be quasi-static and Eq.~\eqref{eq:Shapeom2} can be solved as 
\begin{equation}
\int \frac{d\phi}{[I_{\mathrm{bias}}(\tilde{t} R^{-1}\Omega_J)-I_J(\phi)]}=\int dt.
\end{equation}
Here, we focus on the case where $\tilde{t}$ is such that $I_{\mathrm{bias}}(\tilde{t}R^{-1}\Omega_J) \neq I_J (\phi(\tilde{t}))$ and the phase variable evolves rapidly compared to $I_{\mathrm{bias}} (tR^{-1}\Omega_J)$.
As $\phi$ changes because of the periodic dependence of $I_J$, one must approach
 a minimum of the washboard potential when $[I_{\mathrm{bias}}(\tilde{t} R^{-1}\Omega_J)-I_J(\phi)]\sim 0$ (where the dynamics slows down and the integral on the LHS diverges). There are two relevant time intervals: (i) where 
$I_{\mathrm{bias}}(tR^{-1}\Omega_J)=I_J(\phi)$ has a solution $\phi = \phi_0(t)$ and (ii) where there is no such solution 
(or local minimum of $U_{\mathrm{wb}}$).  

Before analyzing region (i), which will be the focus of our analysis, let us first show that the time range (ii) 
is small. Scaling $t\rightarrow \Omega_J^{-1}t$, the equation of motion [Eq.~\eqref{Shapeom1}] becomes  
\begin{equation}
\frac{d\phi}{dt}=(\Omega_J^{-1} R)[I_{\mathrm{bias}}(t)-I_J(\phi(t))].
\end{equation}
In the limit $\Omega_J^{-1} R\rightarrow\infty$, the phase $\phi$ can change by a period in a 
parametrically small time. Changes by a large number of periods would correspond to a large phase. 
This would correspond to high Shapiro steps as a function of the bias dc current. Therefore, we assume 
that the dc part of $I_{\mathrm{bias}}$ is small enough past the first Shapiro step, so that the time range (ii)
is small. This is the assumption referred to below Eq.~\ref{Shapeom1}.

\begin{figure}
\capstart
\centering
\includegraphics[width=\linewidth]{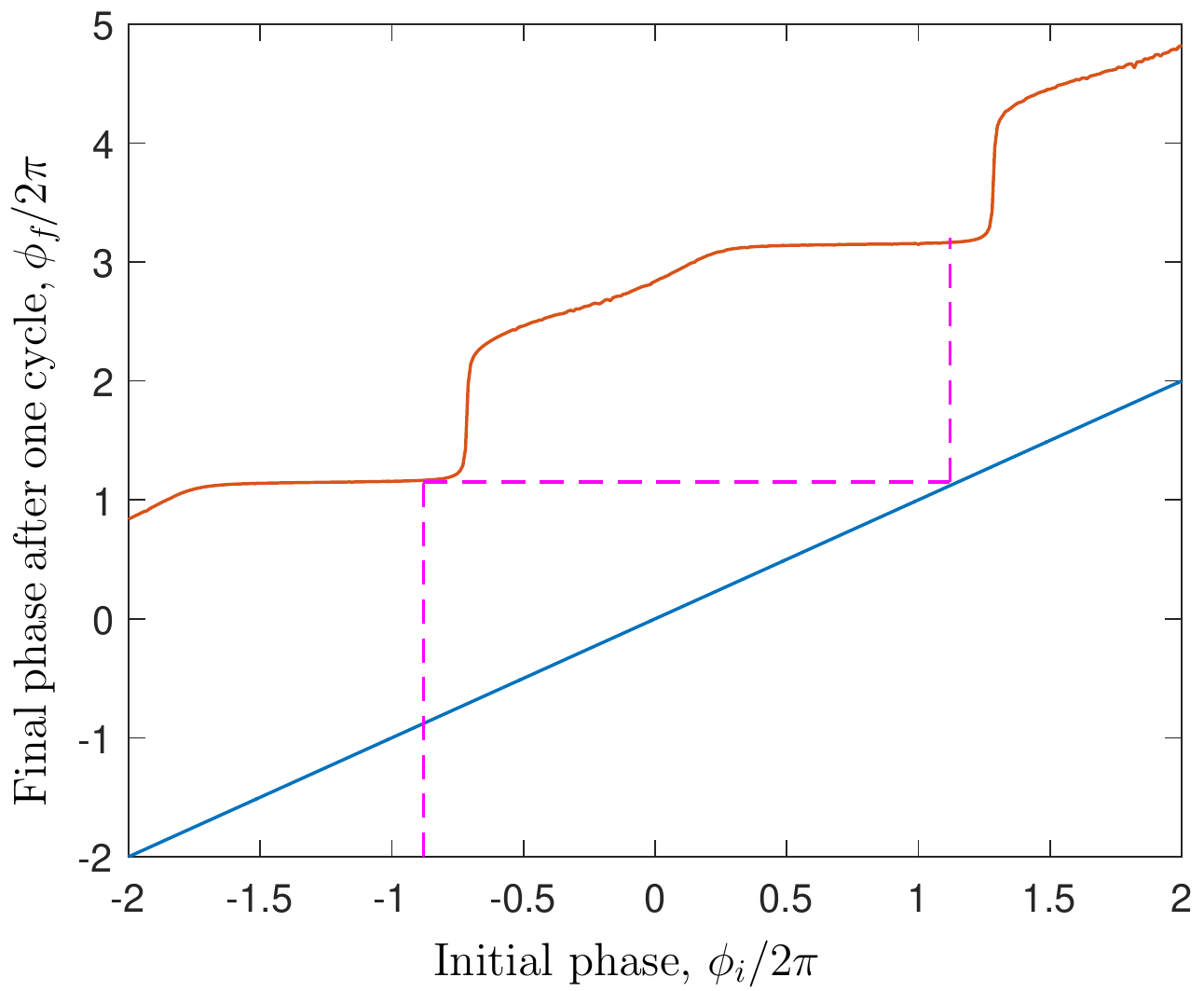}
\caption{(Color online) 
Poincare map for the dynamics of Eq.~\ref{Shapeom1} showing the resulting phase $\phi_f$ after one period given the 
initial phase $\phi_i$. The long-time periodic dynamics can be computed from the Poincare map. As discussed in the main text, the Poincare map shows doubled steps that correspond to the doubled Shapiro step 
seen numerically in Fig.~\ref{Fig3}. 
}\label{supplfig:fig1}
\end{figure}

Under this assumption, the dynamics in region (i) spans most of the 
time. However, based on a similar argument in the previous paragraph, we can argue that the phase dynamics 
is fast when $I_J(\phi)$ is significantly different from $I_{\mathrm{bias}}(t)$. Defining $\phi_0(t)$ in the region (i)
so that 
\begin{equation}
I_J(\phi_0(t)) = I_{\mathrm{bias}}(t),
\end{equation}
we can assume that $\phi(t)$ rapidly evolves until $\phi(t)\sim \phi_0(t)$ (i.e., 
the phase variable approaches a local extremum) where it slows down. However, the dynamics of the phase variable in 
this region can be described by linearization by defining 
\begin{equation}
\delta\phi=\phi-\phi_0,
\end{equation}
whose dynamics is given by the equation 
\begin{equation}
\dot{\delta\phi}+\Omega_J\dot{\phi}_0=-RI_J'(\phi_0)\delta\phi.
\end{equation}
The solution of this equation is written as 
\begin{equation}
\delta\phi(t)=e^{-\Lambda(t)}\delta\phi(0)-\Omega_J\int dt' e^{-(\Lambda(t)-\Lambda(t'))}\dot{\phi}_0(t'),\label{linearized1}
\end{equation}
where 
\begin{equation}
\Lambda(t)=\frac{R}{\Omega_J}\int_0^t dt' I_J'(\phi_0(t'))
\end{equation}
is the Lyapunov exponent of the dynamics. Here $t=0$ represents the time when a particular trajectory approaches 
close to the minimum $\phi_0(t)$. 

Let us now use the picture above to construct the Poincare map of the periodic dynamics shown in Fig.~\ref{supplfig:fig1}.
The Poincare map for a time-periodic system is defined as a function for the phase variable at the end of a 
period $\phi=\phi_f$ in terms of the initial condition $\phi=\phi_i$ at the beginning. Given this function, one can construct the long 
term dynamics of the equation.  Based on the previous paragraph, it is convenient to choose the period at 
the end of the region (i) where $U_{\mathrm{wb}}$ still has a local minimum and the phase particle is converging to the minimum 
because of the negative Lyapunov exponent. Assuming the Lyapunov exponent is large (i.e., $\Omega_J\rightarrow 0$), 
the trajectories of $\phi(t)$ over almost the entire range of $\phi$ at the initial point of region (i) (which we called $t=0$ 
before)
 converge to one of the minima where $\phi\sim \phi_0(t)$. There are, however, some small range of "transition" values of 
$\phi$ at $t=0$ where the trajectories do not approach a minimum.  Apart from this transition region, the rest of the range of 
$\phi$ at $t=0$ is compressed to an exponentially small range in $\phi_f$.
A subtle point to note is that the beginning of region (i) is preceeded by a small range of region (ii) over which the
Lyapunov exponent contribution $I_J'$ is not necessarily positive. This region is the key in connecting the initial time where 
$\phi_i$ is set at the beginning of the period to the time $t=0$ which is the beginning of region (i). It is possible, in principle,
that the range away from the transition region which is compressed to an exponentially small part of $\phi_f$ is generated 
from an exponentially small part of $\phi_i$. However, because the range of time in (ii) is assumed to be parametrically 
smaller than (i), the amplification in region (ii) from $\phi_i$ to $t=0$ is much smaller than the total Lyapunov exponent 
$e^{-\Lambda(t)}$ accumulated over region (i). Therefore, we expect plateaus in $\phi_f$ as a function of the 
initial condition $\phi_i$ as seen in the Poincare map in Fig.~\ref{supplfig:fig1}.

One can determine from the Poincare map in Fig.~\ref{supplfig:fig1} that the long-term dynamics will be characterized by 
a stable attractor where the phase changes by an integer multiple of $4\pi$ over each cycle. To see this, we note that 
for certain values of the dc bias current $I_{\mathrm{dc}}$, the plateau value of the phase $\phi_f$ will occur in a range of $\phi_i$
where the plateau is stable.  This leads to the phase particle returning to the plateau at regular intervals leading to the Shapiro 
step in Fig.~\ref{Fig3}(b). The stability of the trajectory can be further understood by considering 
the Lyapunov exponent around the proposed trajectory $\phi_1(t)$ corresponding to Fig.~\ref{Fig3}(a). By linearizing Eq.~\ref{Shapeom1} similar to 
Eq.~\ref{linearized1}, we see that a solution to Eq.~\ref{Shapeom1} is written as 
\begin{align}
\phi(t)\approx&~\phi_1(t)\nonumber\\
&+[\phi(0)-\phi_1(0)]\mathrm{exp}\left[-\int d\phi \frac{I_J'(\phi)}{I_{\mathrm{bias}}(t_1(\phi))-I_J(\phi)}\right],
\end{align}
where $t_1(\phi)$ is the inverse function of $\phi_1(t)$. Furthermore, we observe that  the integral is dominated by the range of time when the potential has a minimum (as we noted before).
In this case, the denominator of the 
exponential is vanishingly small and dominates the exponent (as we saw for a single period). 
As a result, the Lyapunov exponent for trajectories that approach the minimum remains negative even over the entire 
time period.

\end{document}